# MULTISCALE SIMULATION OF SURFACE NANOSTRUCTURE EFFECT ON BUBBLE NUCLEATION

**Yijin Mao, Bo Zhang, Chung-Lung Chen, Yuwen Zhang**

Department of Mechanical and Aerospace Engineering
University of Missouri
Columbia, Missouri, 65211
Email: zhangyu@missouri.edu

**ABSTRACT**

Effects of nanostructured defects of copper solid surface on the bubble growth in liquid argon have been investigated through a hybrid atomistic-continuum method. The same solid surfaces with five different nanostructures, namely, wedge defect, deep rectangular defect (R-I), shallow rectangular defect (R-II), small rectangular defect (R-III) and no defect, have been modeled at molecular level. The liquid argon is placed on top of the hot solid copper with superheat of 30 K after equilibration is achieved with CFD-MD coupled simulation. Phase change of argon on five nanostructures has been observed and analyzed accordingly. The results showed that the solid surface with wedge defect tends to induce a nano-bubble relatively more easily than the others, and the larger the size of the defect is the easier the bubble generate.

**INTRODUCTION**

In the past few years, the study of nanostructured effect on heat transfer enhancement has already drawn great attentions because of its potentially wide applications in a many fields, such as small device cooling technology [1-5], water treatment [6], and nanomaterials fabrication [7-9]. A large number of research papers can be found in the literature. Lee, et al. [10] compared nucleate pool boiling heat transfer coefficients between non-coated surface and nano-porous coated surface. It was found that the incipient wall superheat of pool boiling in nano-porous coated surface was lower than the other one. The corresponding heat transfer coefficient with nano-porous coated surface appeared higher than that of non-coated surface. Li et al. [1] has reported that nanostructured copper interface demonstrated enhanced boiling performance at low superheated temperature in comparison with untreated surface. It is found that the large increase in the density of active bubble nucleation sites explain this performance enhancement. Forrest et al. [11] also demonstrated nanoparticle thin-film coated surface had a significant enhancement in the pool boiling critical heat flux and nucleate boiling heat transfer coefficient. Chu et al. [12] conducted an experiment to investigate the surface roughness-augmented wettability on critical heat flux during pool boiling with horizontally oriented surface. Their experimental data supported the conclusion that the roughness-amplified capillary forces played an important role in critical heat flux enhancement.

Though many experimental studies have already confirmed that the nanostructured surface could significantly improve the heat transfer performance and critical heat flux, the mechanism behind remain unknown since there is no cost-effective way to observe the incipient sight of nano-bubble generation in terms of the state-of-the-art technology. With the knowledge of bubble formation mechanism, it will definitely be helpful to understand nanoscale boiling precisely, pave the pathway to explore deeper into the field where up-to-date experimental technology cannot reach and certainly help to facilitate an optimized nanostructure for specific engineering purpose.

To achieve this objective, a hybrid atomistic-continuum simulation (HAC) method for two-phase flow has to be developed. It is worth providing a brief background on this methodology. The HAC usually refers to methodologies coupling between molecule dynamics and computational fluid dynamics (MD-CFD), direct simulation Monte Carlo and computational fluid dynamics (DSMC-CFD), and molecular dynamics and finite element method (MD-FEM). The DSMC-CFD can only be feasible to the simulation of dilute gas flow, while MD-FEM usually is only restricted to the solid-state problems. Among all, the MD-CFD hybrid method for dense fluids are probably the most challenging method due to its difficulties in applying the boundary condition onto the MD domain [13]. In addition, within the framework of MD-CFD, conservations of mass, momentum and energy have to be resolved at the same time scale, which means that the boundary conditions for all equations have to be appropriate addressed. Moreover, HAC usually can also be divided into two genres, including state-coupling and flux-coupling [14], according to the manner of data exchanged at hybrid solution interface (HSI). The former one, as indicated by its name, expects to exchange variables of states, such as density, pressure, velocity, and temperature between the two domains, while the latter exchanges mass flux, momentum flux and energy flux through the HSI instead. In addition, the mixing scheme of state and flux coupling can also be found in the literature, which has MD to CFD via state coupling, while CFD to MD through flux coupling. It is well known that the flux coupling scheme is considered to be a good method because the derivation of the flux coupling scheme directly guarantee the conservation of mass, momentum and energy. Nevertheless, it should also be pointed out that the state-coupling scheme does introduce less fluctuation which significantly stabilizes the computation in continuum domain, while the flux coupling does enforce the time step of continuum domain to be same as that of the molecular domain (state coupling assume the time is decoupled by presumably decomposing the transient problem into a series of steady state solutions [15, 16]). In fact, on which type of quantity the coupling should be based on is still a subject of some debates [13].

Nevertheless, such hybrid scheme is still considered as very powerful and accurate, since the molecular domain is able to produce high-resolution data which is essential to affect the continuum domain where sub-models may lack of accuracy. That is why the HAC for micro/nanoscale problems through classical molecular dynamics (MD) and mesh-based methods (FVM, FEM etc.) has received intensive attention in the past two decades [17]; a number of scientific investigations have been conducted to unveil the unknown phenomena that conventional theory cannot explain well, especially those the interface related issues, such as thermal resistance, boiling, and wave propagation etc. [18-23]. However, it has to be admitted that, as a literature survey indicated, most of this kind of numerical simulations have focused on solving steady-state problems. Scant efforts have been placed on the transient problems, not to mention the multiphase problems where phase changes occur, such as boiling, with the fact that MD-CFD for dense fluid itself already is very difficult to solve.

In this work, a serial numerical simulation of nano-bubble growth on nanostructured solid surface are conducted through a hybrid atomistic-continuum simulation where the full state-coupling scheme is adopted to exchange the data at HSI, and a fade-in pairwise potential is utilized to assist the random molecule insertion in the course of density coupling. Certain boundary conditions have been developed to accomplish the data exchange between MD and CFD domains.

**NOMENCLATURE**

| | |
|---|---|
| c | specific heat, kJ/kg K |
| f | force, N |
| g | gravity, m/s$^2$ |
| I | unit tensor |
| k | thermal conductivity, W/m K |
| L | latent heat, kJ/kg |
| m | mass, kg |
| p | pressure, Pa |
| r | position vector, m |
| t | time, s |
| T | temperature, K |
| U | potential energy, J |
| U | velocity, m/s |

**Greek Symbols**

| | |
|---|---|
| $\mu$ | viscosity, Pa s |
| $\rho$ | density, kg/m$^3$ |
| $\sigma$ | surface tension, N/m |
| $\tau$ | relaxation time, s |
| $\omega$ | fading factor |

**Subscript**

| | |
|---|---|
| 0 | start time |
| 1 | liquid phase |
| 2 | vapor phase |
| i | atom index |
| K | curvature |
| l | liquid phase |
| sat | saturation |
| $\rho gh$ | hydrostatic pressure, kg/m3 |
| T | time |
| v | vapor phase |

**Physical Model and Methods**

For the problem of interest, the whole simulation domain is decomposed into three major parts, including continuum domain, molecular domain and hybrid solution interface. The following contents will be introducing the physical model and methods adopted for each of them, respectively.

**Continuum domain**

Since both liquid and vapor phase will be considered in the continuum domain, the governing equations that will be solved include conservation of mass, momentum, energy, along with advection of volume of fluid ($\alpha$), with the concept of volume of fluid methodology (VOF). In addition, an appropriate phase change model also has to be employed for this domain.

With the incompressible assumption for the liquid and vapor phase, the mass conservation can be expressed as,

$$\int_{dV} (\nabla \cdot \mathbf{U}) dV = \dot{m}_1 \left( \frac{1}{\rho_1} - \frac{1}{\rho_2} \right) = \left( \dot{m}_{1,v} + \dot{m}_{1,c} \right) \left( \frac{1}{\rho_1} - \frac{1}{\rho_2} \right) \quad (1)$$

where $\dot{m}_1$ denote the mass change rate of phase 1 which is liquid, while $\dot{m}_{1,v}$ (-) and $\dot{m}_{1,c}$ (+) represent the mass change rate due to vaporization and condensation of liquid, respectively, and $\rho_1$ and $\rho_2$ are density of liquid and vapor.

The momentum equation can be expressed as,

$$\frac{\partial(\rho \mathbf{U})}{\partial t} + \nabla \cdot (\rho \mathbf{U} \otimes \mathbf{U})$$
$$= \nabla \cdot (\mu \nabla \mathbf{U}) + \nabla \cdot \left( \mu (\nabla \mathbf{U})^T - \mu \frac{2}{3} trace(\nabla \mathbf{U})^T \mathbf{I} \right) + \sigma_K \nabla \alpha_1 - \nabla p_{\rho gh} - \mathbf{g} \cdot \mathbf{r} \nabla \rho$$
(2)

where bulk density and viscosity are obtained in the manner of phase weighted average, namely, $\rho = \alpha_1 \rho_1 + \alpha_2 \rho_2$ and $\mu = \alpha_1 \mu_1 + \alpha_2 \mu_2$, and $\sigma_K$ represent the product of surface tension and interface curvature.

The energy equation can be expressed as,

$$\frac{\partial(\rho c T)}{\partial t} + \nabla \cdot (\rho c \mathbf{U} T) = \nabla \cdot (k \nabla T)$$
$$+ \left[ T_{sat} \left( \frac{\partial(\rho c_v)}{\partial t} + \nabla \cdot (\rho c_v \mathbf{U}) \right) - L_v \left( \frac{\partial(\rho_v \alpha_2)}{\partial t} + \nabla \cdot (\rho_v \alpha_2 \mathbf{U}) \right) \right]$$
(3)

where specific heat $c$ and thermal conductivity $k$ are also based on phase weighted average, $T_{sat}$ is saturation temperature, and $L_v$ is latent heat of vaporization, subscript $v$ indicates vapor phase.

Advection of volume of fluid (α) can be expressed as,

$$\frac{\partial \alpha_1}{\partial t} + \nabla \cdot (\alpha_1 \mathbf{U}) - \alpha_1 \nabla \cdot \mathbf{U}$$
$$= \dot{m}_{1,v} \left[ \frac{1}{\rho_1} - \alpha_1 \left( \frac{1}{\rho_1} - \frac{1}{\rho_2} \right) \right] + \dot{m}_{1,c} \left[ \frac{1}{\rho_1} - \alpha_1 \left( \frac{1}{\rho_1} - \frac{1}{\rho_2} \right) \right]$$
(4)

These algorithm of solving these four governing equations are fully implemented in the solver of *interPhaseChangeFoam* which comes with OpenFOAM-2.3.0 [24].

Regarding with the phase change model, it is true that a number of phase change models dealing with boiling or cavitation can be found in the literature [25]. Lee Model [26] considered the phase change rate as a dependency of temperature deviation from the saturation temperature. Sun et al. [27] took the phase change rate as temperature gradient dependent. Kunz Model [28] treated the phase change rate as pressure dependent. In this work, the phase change model that introduced by Sun et al. [27] will be adopted since it was rigorously derived through energy conservation at the liquid-vapor interface.

**Molecular domain**

For any classical molecular dynamics simulation, the trajectory of any single atom will be explicitly found through following equation,

$$m_i \frac{d^2 \mathbf{r}_i}{dt^2} = -\sum_{i \neq j} \mathbf{f}_{ij} + \mathbf{f}_i^{ext}$$
(1)

where the force $\mathbf{f}_{ij}$ between atom i and j will be found through pre-defined potentials and $\mathbf{f}_i^{ext}$ represent the external force caused due to other effect. The evolution of the entire atomic system then can be simulated through integration scheme, such as Verlet.

The open-source molecular dynamic code LAMMPS [29] will be employed for this part. Since this domain is completely independent of the continuum domain, not only the monoatomic system with pairwise potential can be supported, but the polyatomic system with various potentials, such as Reaction Empirical Bond Order (REBO), are also fully supported as long as it is compatible with LAMMPS.

**Hybrid solution interface (HSI)**

Without any doubt, the most important part of any HAC methodology is the way to synchronize the variables of two domains. As mentioned previously, there are two types of coupling scheme in the literature and each one of them has its own deficit/merit over the other.

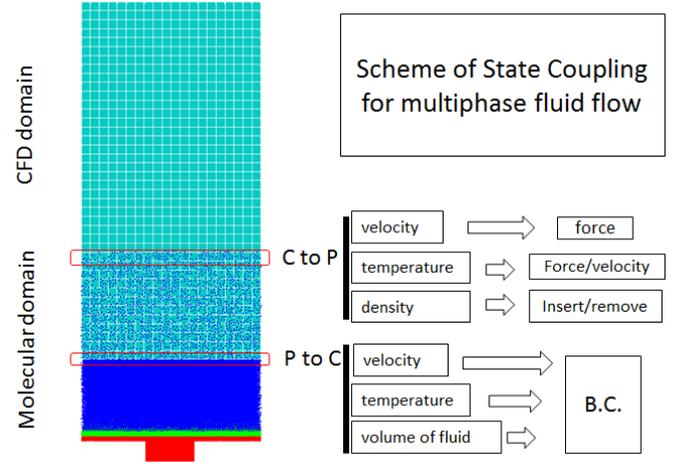

**Figure 1** Scheme of State Coupling for Multiphase fluid flow

As presented in the Figure 1, the hybrid solution interface encompass *C to P* and *P to C* interfaces. The former (*C to P*) refers to the data transfer from the continuum domain to the molecular domain (or particle domain), while the latter (*P to C*) represents the data transfer from the molecular domain (or particle domain) to the continuum domain. Other than velocity and temperature, the third state variable, the volume of fluid (α), has been included for the *P to C* interface for this liquid-vapor system. In the meantime, the density match is considered at the interface of *C to P* due to advection of volume of fluid (α) in the continuum domain. The implementations applied into the C to P region and P to C region are described below.

**C to P region**

Because the velocity obtained from the continuum domain should be always consistent with the one from MD domain, it implies that the mean momentum from the MD domain should be equal to the instantaneous macroscopic momentum from the continuum domain. In order to achieve this momentum consistency, it is better to provide an external force that is proportional to the momentum difference between CFD-grid given value and current MD given value. For the temperature consistency, another temperature-dependent force should be applied. In fact, there are other thermostat existing in literature;

for example, directly scaling the velocity of each atom to the desired temperature. More detail on how to compute these forces could be found in our previous works [23, 30-32].

To the multiphase system, the density of the fluid is changeable due to advection of phases; therefore, an artificial operation of inserting/removing molecules should be adopted to meet the density consistency at *C to P* interface. The previous studies [14, 33] does provide a candidate method inserting/removing molecules according to the mass exchange. In this work, the number of atom (or molecule) is directly controlled in according to the density of the given CFD cell, as shown in Figure 2. During the process, the inserted atoms (or molecules) are randomly placed into the molecular domain, with a minimum distance from the existing ones, rather than using USHER [34, 35] algorithm which may not be sufficiently general to be implemented into the LAMMPS code.

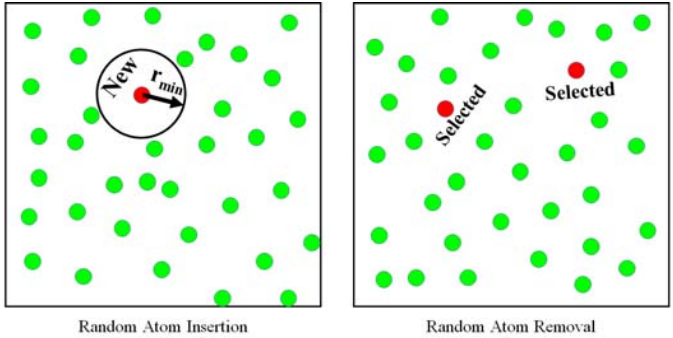

**Figure 2** Demonstration of molecule insertion/removal.

In addition, to ensure the computational stability, a Fade-In type potential function [36] where a time-dependent coefficient is introduced to relax the possible abrupt energy state of the newly inserted molecules. The mathematical form of the atomic force and coefficient $\omega_{ij}$ can be expressed as,

$$\mathbf{f}_{ij} = -\omega_{ij} \nabla U \quad (2)$$

$$\omega_{ij} = \begin{cases} \left[2(t-t_0)/\tau_T\right]^n / 2, & \text{if } t < \tau_T/2 \\ 1 - \left|\left(2(t-t_0-\tau_T)/\tau_T\right)^n\right|/2, & \text{if } \tau_T/2 < t < \tau_T \\ 1, & \text{if } t > \tau_T \end{cases} \quad (3)$$

where $\tau_T$, $n$ and $t_0$ are relaxation time, power, and the time when new molecule is created.

**P to C region**

In order to compute the velocity, temperature and alpha (α) values at the cells immersed in the molecular domain, an invisible box with its origin located at the center of the boundary face for each cell is created, as shown in the Figure 3. The state values within the box will be computed in a time average manner and then transferred back to the *P to C* interface.

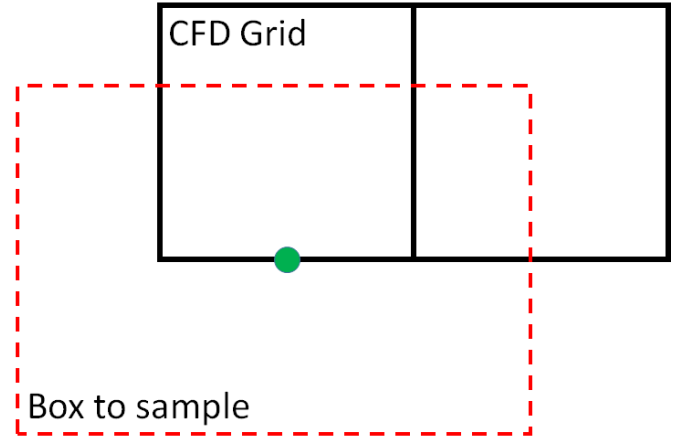

**Figure 3** Illustration of the box for P to C data transfer

**Model Demonstration**

The nanoscale-boiling problem itself is a transient problem; therefore, whether the molecular domain is capable to inspect the information carried by CFD-grids quickly and accurately is a critical issue that should be carefully examined. In order to verify if the present scheme is capable to alter the behavior of molecules in such a fashion, a case that has *C to P* cells been applied with a sinusoidal-like velocity, temperature, and density is prepared for the validation.

The 387 Å×174 Å×417 Å molecular box is filled with 50,589 Lennard-Jones fluid argon which is equilibrated to temperature of 116 K and 17,120 Lennard-Jones like copper atoms with same temperature. The interaction parameters used are $\varepsilon_{Ar-Ar}$ = 0.2381 Kcal/mole, $\sigma_{Ar-Ar}$ = 3.405 Å, $\varepsilon_{Cu-Cu}$ = 9.566 Kcal/mole, and $\sigma_{Cu-Cu}$ = 2.2277 Å. The interaction between argon and copper comply with the sixth power law [29] and the cutoff is 13 Å. The CFD domain has dimension of 387 Å×17.4 Å×416 Å, with the height of overlap region as 334 Å. The timestep for MD and CFD are 5 fs and 5 ps, respectively. The relaxation time for Fade-in Lennard Jones potential is $\tau_T$=1000 fs and *n*=1.

The MD and CFD data will be exchanged every 1000 MD timesteps. The right side of the Figure 4 gives the variation of state variables within the sample region. As can be seen, all the state variables, which includes velocities, temperature and density, are quickly and accurately synchronized with the state values given by the CFD cells. As it is demonstrated for the velocity coupling, the sinusoidal-like velocity in the sampled region can be effectively prohibited from exceeding ±100 m/s which occur at timestep of 6000, 11000, 16000 and so on (removal/insertion occur), while for the rest timesteps, it is found that there is negligible difference between two simulation domains. In the other words, the current scheme is sufficiently sensitive to recover the states along with that of the continuum domain.

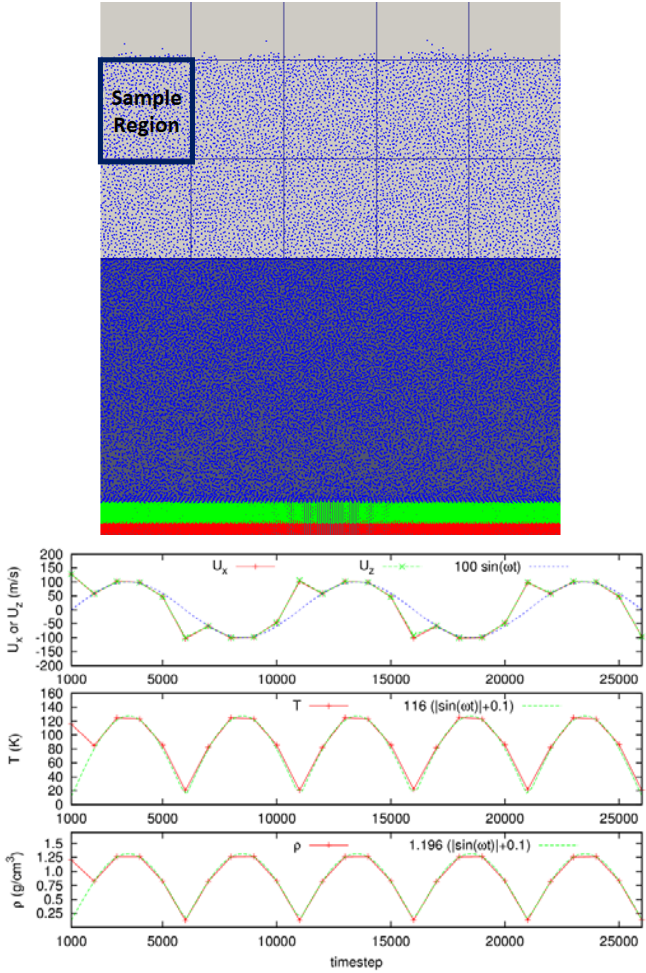

**Figure 4** Velocity, temperature and density profiles at the sampled cell along with the user specified sinusoidal function.

## Results and Discussions

In order to investigate the thermal induced bubble growth at the solid surface with defect, starting from an appropriate fluid state where density difference between liquid and vapor are relatively small would be a good choice, in consideration of maintaining a smaller Knudsen number. In according to data provided by NIST [37], the chosen state of liquid argon has the boiling point of 116 K at pressure of 10 MPa. The corresponding thermophysical properties are also found through NIST database, as given in [37]. The dimension of the MD domain is 566 Å ×13.2 Å ×522Å; while it is 566 Å ×13.2 Å ×2,650 Å for the CFD domain. The CFD domain has mesh configuration of 10×1×53. The timestep for MD and CFD are 5 fs and 5 ps, respectively. A periodic boundary condition is applied to the x- and y- directions, while for the z direction, a boundary is non- periodic. The MD and CFD data will be exchanged every 1000 MD timesteps. A Lennard-Jones wall is placed on top of *C to P* region to enforce a spatial redistribution of the atoms close to the MD boundary. The Lennard-Jones interaction parameters are 3.405 Å and 0.7 eV, respectively. For *P to C* region, the boxes for sampling velocity, temperature and α are cubes with length of 200Å, 200Å and 80Å, respectively. The molecular domain itself is equilibrated for 1 million timesteps before bridged into CFD domain. After that, the whole computational domain is equilibrated for 1.25 ns before the solid is immediately heated up to 146K (superheat 30K). All thermostats used in this work are achieved by rescaling the either absolute or thermal velocity of atoms.

**Table 1** Parameters for continuum domain of argon

|  | liquid | vapor |
|---|---|---|
| ρ | 1196.5 kg/m$^3$ | 49.5kg/m$^3$ |
| ν | 0.0998 mm$^2$/s | 0.206 mm$^2$/s |
| k | 0.087 W/mK | 0.00887 W/mK |
| c | 1.285 KJ/Kg/K | 0.8 KJ/Kg/K |
| σ$_s$ | 0.00567 N/m | |
| L | 112 kJ/kg | |

**(a) Wedge-shaped defect**

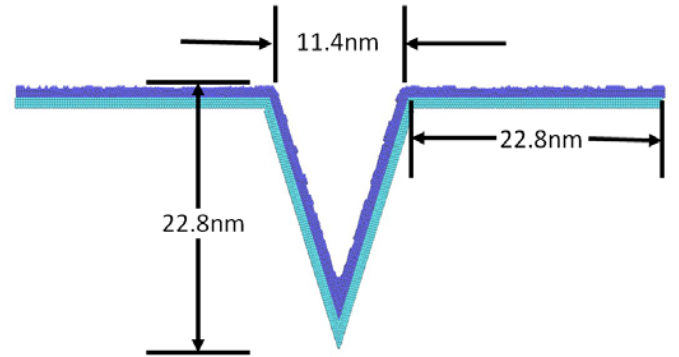

Figure 5 Copper surface with wedge-shaped defect

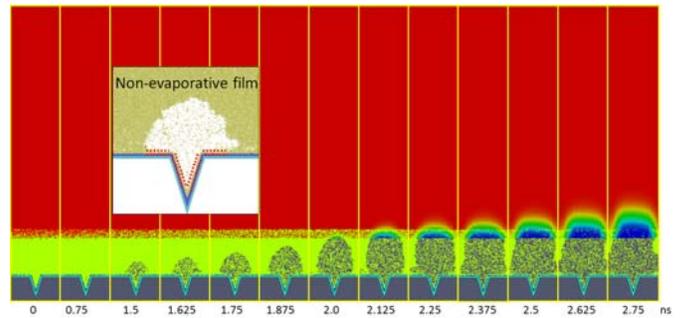

Figure 6 Bubble growth at the solid surface with wedge-shaped defect

Figure 5 shows the geometry and dimensions of the wedge-shaped defect. In order to minimize the computational cost, the solid copper is composed of two cubic type layers. One is stationary wall, the other is mobile atoms where the temperature is scaled to the desired value. Figure 6 gives the process of nano-size bubble growth after the solid copper is immediately heated up to 146K. It is found that a small vaporized-zone emerged in the wedge-shaped defect after 0.75ns heating. It is worth mentioning that the nano-bubble starts interacting with the bottom of CFD domain at time of 2.0 ns and then successfully migrates into the continuum domain and keep expanding naturally in the domain. In addition, a close observation at the solid-vapor interface reveals the existence of a non-evaporative liquid film with a few atomic thickness, as shown between dash lines.

**(b) Rectangle-shaped defect I (deeper, R-I)**

Figure 7 shows the geometry and dimensions of the deeper rectangle-shaped copper surface. As can be seen, the configuration of the solid plate is exactly same to the solid with wedge defect except for the shape. The depth is 22.8 nm and width is 11.4nm, which are same to the heights and width to the previous case. Figure 8 shows the process of nano-bubble growth during the first 2.75ns after the solid is heated upto 146K. As shown, there is a small vaporized-zone in the defect at time of 0.75 ns, which is similar to the previous case. Later on, the bubble continues growing and at time of 2.5 ns starts to expand into the CFD domain naturally. Similar to the previous one, a non-evaporative liquid film is always placed between vapor and solid interface.

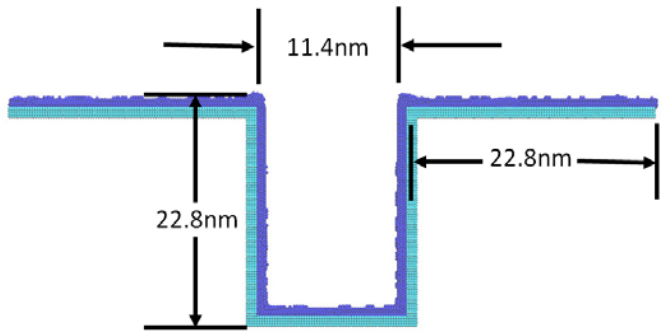

**Figure 7** Copper surface with deeper rectangular defect

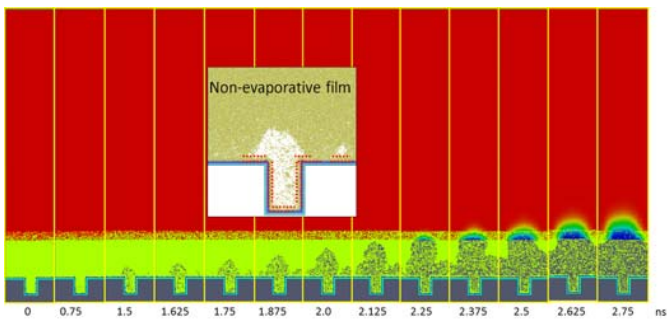

**Figure 8** The process of bubble growth at the solid surface with a deeper defect

**(c) Rectangle shaped defect II (shallow, R-II)**

Figure 9 shows the structure of the solid surface, which has 11.4 nm in both depth and width for the defect. Again, once the solid is immediately heated to 146 K, there is a small vaporized-zone emerged at time of 0.75 ns before which no observable bubble appears, as shown in Figure 10. Interestingly, a smaller bubble also is induced nearby the defect during the process of the bubble growth at time of 1.626 ns. The two bubbles start emerging at time of 2.125 ns and finally become a larger bubble at time of 2.25 ns. Later on, the whole bubble continues expanding and migrates into the CFD domain at time of 2.5 ns. Again, the non-evaporative thin argon liquid film exist between vapor argon and solid copper during the nucleation.

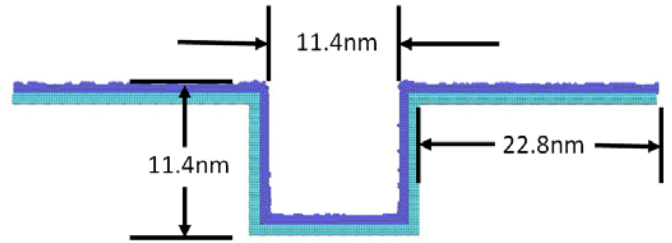

**Figure 9** Copper surface with shallow rectangular defect

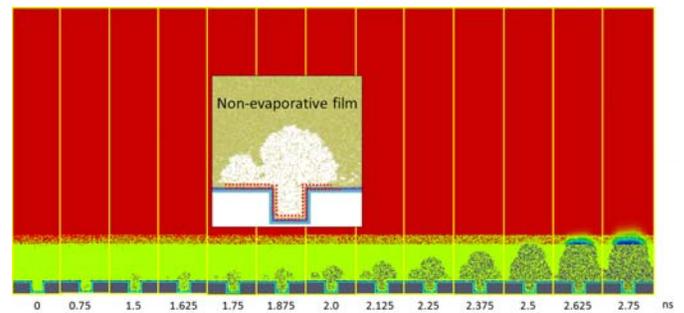

**Figure 10** Bubble growth observed on the surface with shallow rectangular defect

**(d) Rectangle shaped defect III (small, R-III)**

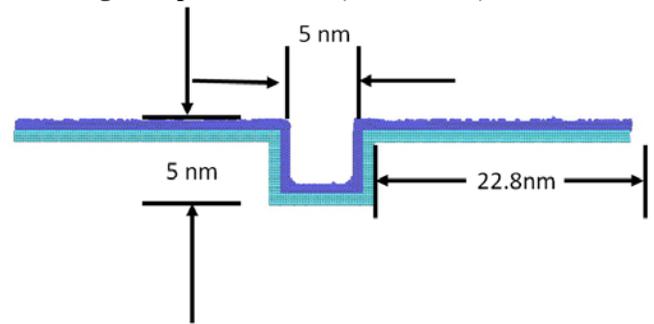

**Figure 11** Copper surface with small rectangular defect

Figure 11 shows the solid surface with smaller defect, which has dimension of 5 nm×5 nm. As indicated in Figure 12, the phenomena observed is significantly different from the previous cases where the sizes of defect are about 4 times larger. In fact, there are a few observable small vaporized-zones appear during the first 2.75 ns. Interestingly, the vapor is not necessarily close to the defect, as shown in the snapshots at time of 2.0 ns and 2.125 ns. However, it seems the defect, though smaller, still has a positive effect on inducing a nano-bubble as demonstrated during the period from 2.375 ns to 2.75 ns. Figure 13 gives a close observation at the solid-liquid interface during the heating process. As circled in the figure, the vaporized-zone tends to form at the zone nearby the cavity (c), however, it will annihilate during the interaction with its surrounding cold liquid (f).

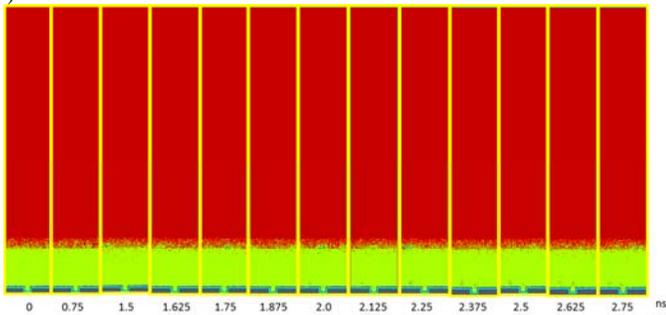

**Figure 12** Only small densify fluctuation exist on the surface with small defect

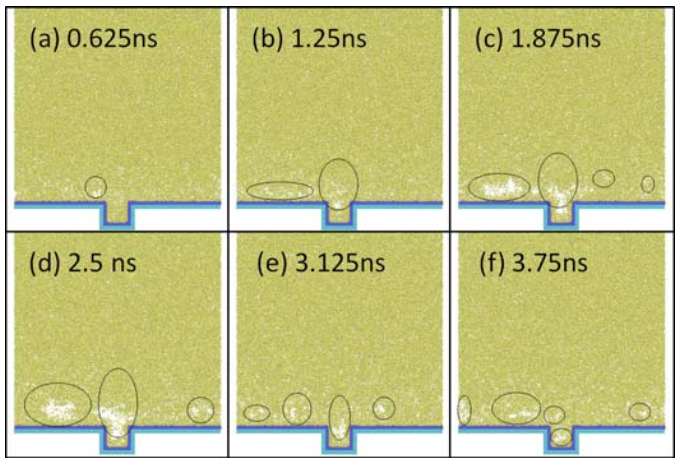

**Figure 13** Density fluctuation at the liquid-solid interface for R-III

**(e) Smooth surface**

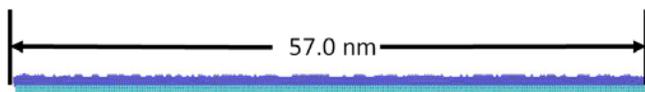

**Figure 14** Copper surface with no defects

Figure 14 shows a 57 nm long solid surface without any defect. The observation once the heat is applied to the solid wall is given in Figure 15. In this case, there is completely no visible vapor appear during the period of 2.75ns. This phenomenon is also significantly different from the previous four cases. Figure 16 shows the local density fluctuation at the liquid-solid interface. As indicated, much less and smaller vaporized zones could be observed in comparison with the case R-III. This fact confirms the positive impact of presence of cavity on bubble nucleation at the solid surface. It could be caused by small energy deviation on the top of smooth solid plate. In fact, the energy deviation is mainly affected by the shape of defects. Therefore, it would be reasonable to speculate that the surface with wedge-shaped defect has the largest energy deviation, while the smooth surface has the minimum.

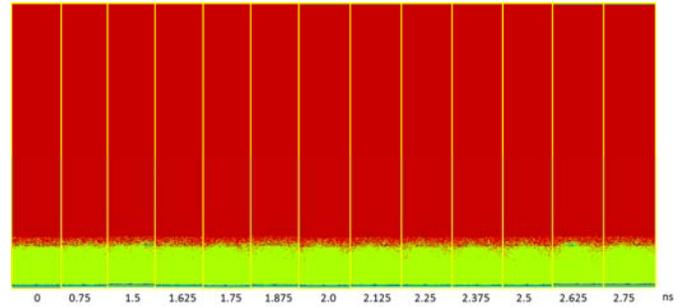

**Figure 15** No vaporization occur during the first 2.75 ns

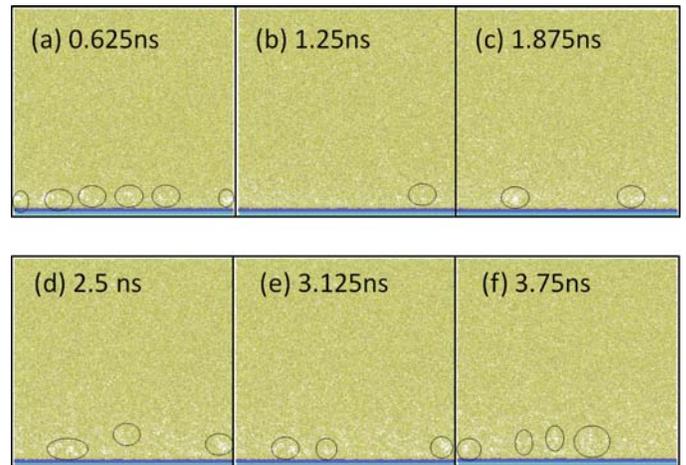

**Figure 16** Density fluctuation at the smooth liquid-solid interface

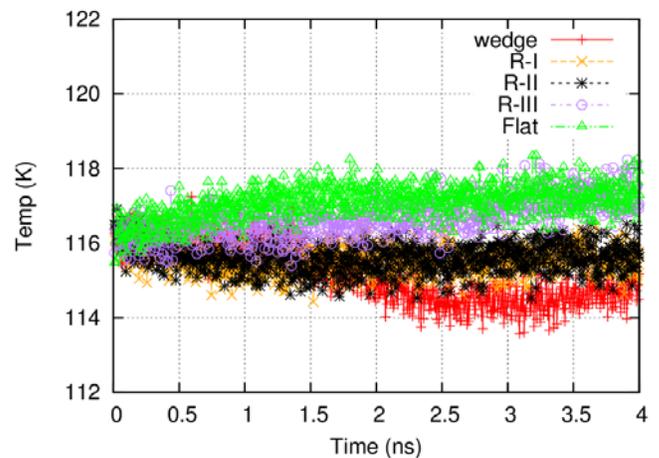

**Figure 17** Temperature variation at a sample region in C to P domain

Figure 17 shows temperature variation at a controlled region that is within C to P domain, as can be seen that the temperature for cases with small rectangle defect (R-III) and without defect (Flat) shows an increasing trend in temperature during heating process, while the other three (Wedge, R-I, and R-II) predict a temperature drop which is caused by energy absorption in the course of vaporization. This phenomena further confirms that the defect size and shape do play an important role in the bubble nucleation of argon liquid on copper solid.

Figure 18 shows the variation of atom number in molecular domain during the entire heating process. As can be seen, the line for the case with wedge-shape defect starts to sharply decline first (time = 1.1ns), the line for the case with rectangle-shape defect I (R-I) decrease later (time = 1.25ns), the one for the case with rectangle-shape defect II (R-II) follows then (time = 1.5ns), and the ones for the case with rectangle-shape defect III (R-III) and without defect (flat) do not show any abrupt decrease in atom number though they are declining. In terms of the slopes of the first three curves, it can be found that the phase change rate follows in the same order, W > R-I > R-II, as the difficult degree for bubble generation.

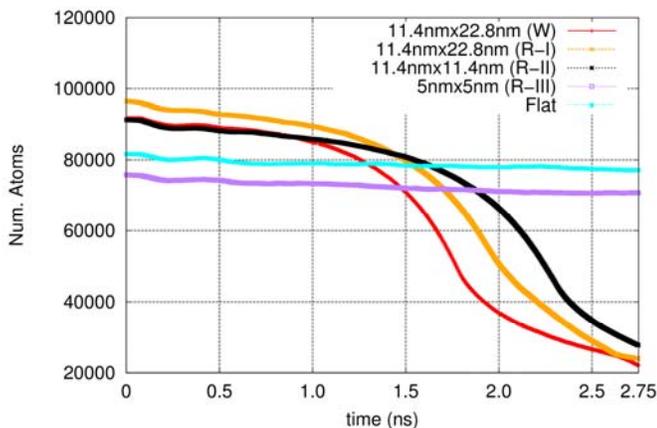

**Figure 18** The number variation of atoms in molecular domain during the first 2.75ns

**Conclusions**

A hybrid atomistic-continuum method for solving bubble nucleation has been developed. This approach is utilized to study the nanostructured effect on bubble generation on defected solid copper surface during the heating process. Surface structures with five different defects, including wedge-shaped defect, rectangle-shaped I (R-I), rectangle-shaped II (R-II), rectangle shaped III (R-III), are investigated.

It is found that wedge shaped surface, though the surface area in contact is not the largest, tends to induce a nano-bubble easier than the rest; while the rectangle-shaped structure (R-I) with deeper cavity shows the second, the other rectangle-shaped structure (R-II) with shallow depth also induce a nano-bubble but later than the deeper one. A non-evaporative film could always be observed between vapor argon and solid copper during the entire process. It is also observed that the square shaped cavity with smaller size (5nm×5nm) seems difficult to induce a nano-bubble in comparing with that of larger defect (11.4nm×11.4nm).

**Acknowledgements**
The authors greatly appreciate the support of the Office of Naval Research under Grant No. N00014-14-1-0402.